\documentclass[a4paper, twocolumn,10pt, unpublished]{quantumarticle}
\pdfoutput=1
\usepackage[utf8]{inputenc}
\usepackage[T1]{fontenc}
\usepackage{amsmath,amssymb}
\usepackage{graphicx}
\usepackage{booktabs}
\usepackage[numbers]{natbib}
\usepackage{hyperref}
\usepackage{preamble/quchip}

\begin{document}
\title{quchip: A Differentiable Toolkit for Modeling Quantum Devices}

\author{Ibraheem AlYousef}
\affiliation{King Fahd University of Petroleum \& Minerals,
Dhahran 31261, Saudi Arabia}
\homepage{https://iPhysics.sa}
\orcid{0000-0002-9467-0616}
\email{Ibraheem@iPhysics.sa}

\begin{abstract}
Predictive modeling of a superconducting quantum chip requires more than a
Hamiltonian: the model must connect device physics, control-line
transformations, chosen frames and approximations, dissipation, and measured
observables. We present quchip, an open-source Python toolkit that represents
these parts explicitly and assembles backend-independent simulations for
QuTiP or dynamiqs; with dynamiqs, device and control parameters remain
differentiable through the solve. We demonstrate the resulting experimental
loop on a five-device model fitted to dressed observables. Simulated phase
sweeps identify the complex crosstalk between two control lines, and inversion
of the inferred response suppresses the effective leakage by more than two
orders of magnitude. Over sixteen simultaneous $\pi$ pulses, the corrected
pulse-end populations remain within $0.3$ percentage points of the
crosstalk-free response. Adiabatically eliminating the bus and readout
resonators reduces the Hilbert-space dimension from 576 to 16, after which
gradients through simulated tomography recover the four injected crosstalk
parameters with a maximum complex error of $1.5\times10^{-5}$. A single
explicit model can therefore support prediction, correction, and inverse
parameter recovery.
\end{abstract}
\maketitle

\section{Introduction}\label{sec:intro}

Simulating a superconducting quantum chip means stitching together and
solving several Hamiltonians. Here, a \emph{device} is a modeled subsystem
with its own local Hilbert space and Hamiltonian, such as a qubit or
resonator. The chip model combines the local dynamics of these devices,
the interactions between them, the drives and control lines that address
them, and the Lindblad collapse operators that account for dissipation.
The simulation itself rests on a sequence of
choices: a frame of reference, the approximations applied within it, the
truncation of each mode, and the expectation values one must compute.
Once these choices are fixed, the Hamiltonian and collapse operators
follow mechanically. They are nevertheless often assembled by hand,
making those choices harder to inspect and the model harder to extend.

Excellent libraries exist across this stack:
scqubits~\cite{groszkowski2021scqubits} and
SQcircuit~\cite{rajabzadeh2023sqcircuit} construct circuit Hamiltonians
and analyze their static properties, while
QuTiP~\cite{johansson2012qutip,lambert2024qutip5} and
dynamiqs~\cite{guilmin2025dynamiqs} solve time evolution and
dissipation. Qiskit Dynamics~\cite{puzzuoli2023qiskitdynamics} provides
JAX-compatible dynamical models, signals, and transfer functions. C3
integrates device and control-electronics models with calibration and
model learning~\cite{wittler2021c3}, while SuperGrad differentiates
superconducting-processor models for device design and parameter
fitting~\cite{wang2025supergrad}.

\begin{figure*}[t]
  \centering
  \includegraphics[width=0.92\textwidth]{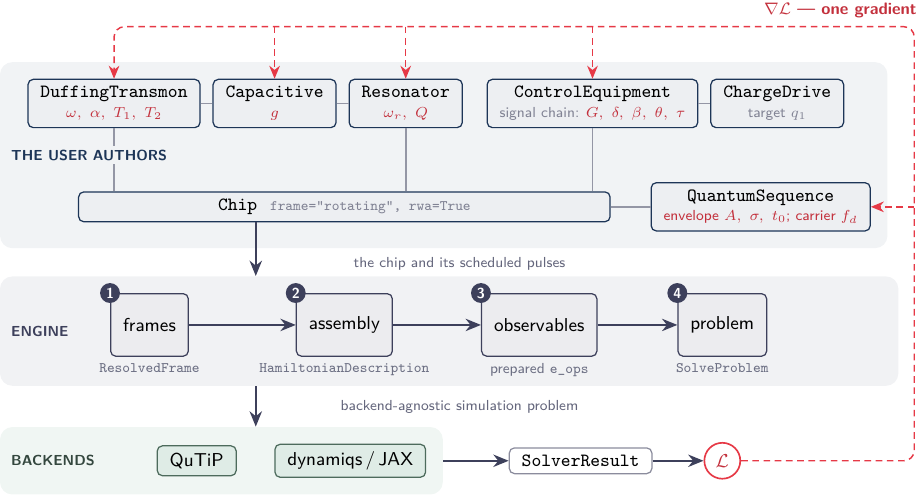}
  \caption{The quchip pipeline. Forward (top to bottom): a four-stage engine
  converts devices, couplings, drives, and the classical signal chain into a
  backend-agnostic physics description, which each backend packs into its
  native solver form. Backward (dashed red): gradients of a scalar loss reach
  device frequencies, coupling strengths, crosstalk entries, carriers, and
  envelope shapes.}
  \label{fig:pipeline}
\end{figure*}

We introduce quchip\footnote{\url{https://github.com/quchip/quchip}}, an
open-source Python package for the physics-to-solver layer between a chip
model and its numerical time evolution. Users describe a chip in terms of
its devices, their couplings, drives, and control lines
(Fig.~\ref{fig:anatomy}). Each declared component contributes a Hamiltonian
or noise term, or a transformation of the classical control signals. From
this description, quchip's engine resolves the reference frame assigned to
each device, applies and records approximations, prepares collapse operators
and observables, and packs a differentiable, backend-independent simulation
problem. QuTiP or dynamiqs then converts that problem to its native solver
form. The resolved frame, band decomposition, observables, and classical
signal path remain explicit through this boundary. The same parameters can
be swept with either backend and, with dynamiqs, remain
JAX-traceable~\cite{bradbury2018jax} through the solve.
Figure~\ref{fig:pipeline} summarizes the forward simulation path and the
differentiable path back to the declared parameters.

This paper makes three contributions:
\begin{enumerate}
  \item \textbf{Explicit, extensible physics.} Devices define their local
  Hamiltonians and noise channels, couplings define interactions, and
  drives define how classical signals enter the Hamiltonian. The engine
  combines these contributions, resolves the reference frame assigned to
  each device, and applies system-wide
  approximations without containing model-specific physics
  (Sections~\ref{sec:building} and~\ref{sec:engine}).

  \item \textbf{An explicit control signal path.} Scheduled pulses produce
  signals that pass through a composable classical signal chain before
  entering the drive Hamiltonians. Gain, delay, and crosstalk are therefore
  represented as properties of the control lines rather than inserted
  directly into the system Hamiltonian. Section~\ref{sec:wiring} defines
  this signal path, and Section~\ref{sec:models} uses it to identify and
  correct control-line crosstalk.

  \item \textbf{End-to-end differentiability.} With the JAX-native dynamiqs
  backend, declared device and control parameters remain traceable through
  signal transformation, Hamiltonian assembly, time evolution, and observable
  extraction. Section~\ref{sec:models} demonstrates this path by inferring the
  four directed crosstalk parameters of the reduced two-qubit model.
\end{enumerate}
We demonstrate all three contributions on an example five-device chip:
Section~\ref{sec:building} constructs its device physics and control
signal path, Section~\ref{sec:engine} follows them through simulation and
differentiation, and Section~\ref{sec:models} identifies and corrects the
declared line crosstalk.

\section{Building a Chip}\label{sec:building}

Researchers describe a quantum chip by the devices it contains, how they
interact, and which control lines drive them. A chip is built in quchip in
the same terms. Each declaration contributes one part of the model. Adding
devices supplies local Hamiltonians and noise channels, adding couplings
supplies interactions, and adding drives specifies how classical line
signals enter the Hamiltonian.

We demonstrate quchip on one running system, the Q1--Q2 subsystem of
the eight-qubit ring on which Balewski et
al.~\cite{balewski2025crosstalk} characterized drive crosstalk: two
fixed-frequency transmons coupled through a shared bus resonator, each with
its own readout resonator~\cite{hashim2025midcircuit}, and driven by two
charge lines that leak onto one another with independent amplitude and
phase in each direction. We construct its Hamiltonian term-by-term: first
the devices (Section~\ref{sec:devices}), then their couplings
(Section~\ref{sec:couplings}), and finally the drives and classical wiring
that carry the leakage (Section~\ref{sec:wiring}).

\subsection{Devices}\label{sec:devices}

A device owns its local Hamiltonian, independent of what it will be
coupled to or driven by. A fixed-frequency transmon, truncated at
quartic order in its Josephson potential, carries the Duffing form:
\begin{equation}\label{eq:duffing}
  H_q = \omega\,\hat n + \tfrac{\alpha}{2}\,\hat n(\hat n - 1)
\end{equation}
with $\omega$ the bare $0 \to 1$ transition frequency and $\alpha$ the
anharmonicity, conventionally negative for a
transmon~\cite{koch2007transmon,krantz2019guide}. Throughout, $\omega$
denotes angular frequency; quchip stores ordinary frequencies
$f=\omega/2\pi$ in GHz and times in ns. quchip implements
Eq.~\eqref{eq:duffing} as \code{DuffingTransmon}:
\begin{lstlisting}
def local_hamiltonian(self, op: LocalOps):
    n = op.n
    return (self.freq * n
        + 0.5 * self.anharmonicity * (n @ (n - op.I)))
\end{lstlisting}
The method matches Eq.~\eqref{eq:duffing} term for term on a Fock space
truncated at \code{levels} (three by default). The class name makes the
approximation explicit; \code{ChargeBasisTransmon} provides the exact
charge-basis model.

The bus and the two readout resonators are instances of a single model:
\begin{equation}\label{eq:resonator}
  H_r = \omega_r\,\hat n
\end{equation}
quchip represents this model with the \code{Resonator} subclass:
\begin{lstlisting}
class Resonator(DeviceModel):
    freq: Scalar = parameter(unit="GHz")

    def local_hamiltonian(self, op: LocalOps):
        return self.freq * op.n
\end{lstlisting}
A new device model follows the same pattern: subclass
\code{DeviceModel}, declare its parameter fields, and implement
\code{local\_hamiltonian}.

The running system's devices are then five declarations:
\begin{lstlisting}
q1  = DuffingTransmon(freq=f1, anharmonicity=alpha)
q2  = DuffingTransmon(freq=f2, anharmonicity=alpha)
bus = Resonator(freq=f_bus, levels=4)
r1, r2 = Resonator(freq=fr1), Resonator(freq=fr2)
\end{lstlisting}
Together they carry the chip's bare Hamiltonian so far:
\begin{equation}\label{eq:htot-devices}
  H_{\mathrm{dev}} = \sum_j \omega_j\, \hat n_j
    + \sum_{k=1}^{2} \tfrac{\alpha}{2}\, \hat n_k(\hat n_k - 1)
\end{equation}
with $j$ running over all five devices and $k$ over the two transmons,
each $\hat n$ acting on its own truncated space.
Dissipation enters either here, as constructor fields (\code{T1=},
\code{T2=}, \code{quality\_factor=}), or at any later point as attribute
writes on the same objects; we defer it to Section~\ref{sec:dissipation}.

\subsection{Couplings}\label{sec:couplings}

A coupling owns the interaction between the two devices it couples, on
their joint space; the devices themselves carry no reference to one
another. The capacitive interaction takes the dipole--dipole form:
\begin{equation}\label{eq:capacitive}
  H_{\mathrm{int}} = g\,(\hat a + \hat a^\dagger)(\hat b + \hat b^\dagger)
\end{equation}
with $g$ the coupling strength in
GHz~\cite{blais2004cavity,blais2021circuit}.
\noindent\begin{minipage}{\columnwidth}
The class \code{Capacitive} declares this form:
\begin{lstlisting}
def interaction(self, a, b):
    return self.g * a.x * b.x   # x = $a + a^\dagger$
\end{lstlisting}
\end{minipage}
A new static coupling follows the same pattern: subclass
\code{CouplingModel} and implement \code{interaction}. The
rotating-wave counterpart of a static coupling is not written by
hand; the engine derives it from the interaction's structure
(Section~\ref{sec:rwa}). Only a time-modulated coupling declares its own
pumped form, since which sideband a pump addresses is physical intent
rather than structure. Whether the RWA is applied is set per coupling
(\code{rwa=}) or for all couplings at once on the \code{Chip}
constructor (Section~\ref{sec:assembly}), with the per-coupling value
taking precedence.

The running model contains four capacitive edges:\footnote{A direct
capacitive edge between the transmons can also be included for a more
complete representation of the device.}
\begin{lstlisting}
couplings = [Capacitive(q1, bus, g=g1b),
             Capacitive(q2, bus, g=g2b),
             Capacitive(q1, r1, g=g1r),
             Capacitive(q2, r2, g=g2r)]
\end{lstlisting}
The four edges complete the static lab-frame Hamiltonian:
\begin{equation}\label{eq:htot-couplings}
\begin{split}
  H_{\mathrm{static}} = H_{\mathrm{dev}} + \sum_{k=1}^{2} \Bigl[\,
    & \cplterm{g_{kb}\,(\hat a_k + \hat a_k^\dagger)(\hat b + \hat b^\dagger)} \\
    &+ \cplterm{g_{kr}\,(\hat a_k + \hat a_k^\dagger)(\hat c_k + \hat c_k^\dagger)} \Bigr]
\end{split}
\end{equation}
with $\hat a_k$, $\hat b$, and $\hat c_k$ the lowering operators of
transmon $k$, the bus, and readout resonator $k$. Every edge enters in
full form.

\begin{figure*}[t]
  \centering
  \includegraphics[width=0.92\textwidth]{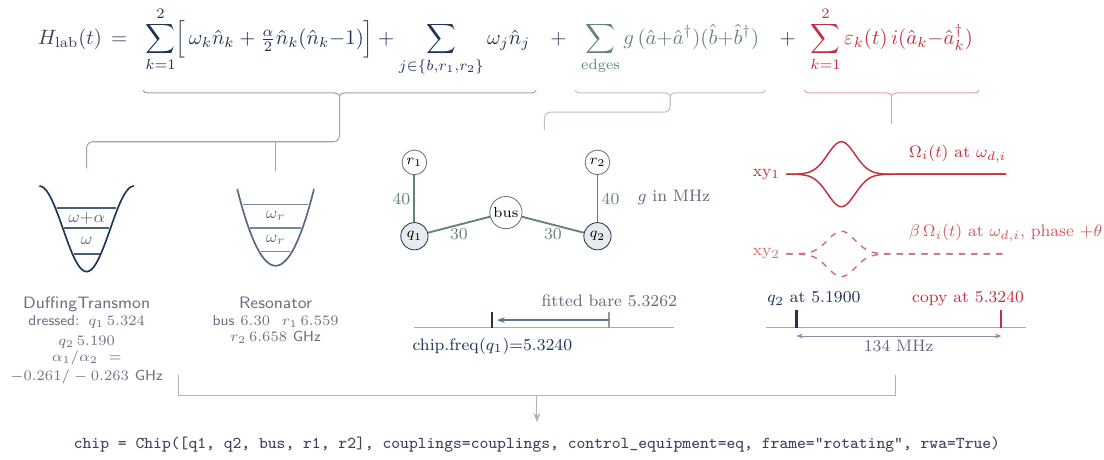}
  \caption{The five-device model used throughout the paper. Each Duffing
  transmon couples to the shared bus and its own readout resonator. Two
  charge lines drive the transmons with crosstalk in both directions; each
  source--victim path has an independent amplitude and phase. Colors match
  the device, coupling, and drive terms in the Hamiltonian.}
  \label{fig:anatomy}
\end{figure*}

\subsection{Drives and Control Wiring}\label{sec:wiring}

A drive specifies how the signal on its line couples to the target
device. A microwave charge line coupled to a transmon addresses the charge
quadrature $i(\hat a-\hat a^\dagger)$~\cite{krantz2019guide}. Its envelope,
drive frequency, and phase are supplied when a pulse is scheduled, so the
declared lines are experiment-independent. The running system has one line
per qubit:
\begin{lstlisting}
xy1 = ChargeDrive(target=q1)
xy2 = ChargeDrive(target=q2)
\end{lstlisting}
A new drive class declares which coupling operator it addresses on its
target (the charge, flux, or phase protocol); scheduling, the signal
chain, and assembly are inherited.

Drive crosstalk sends an RF copy of a source line's output to a victim
line~\cite{sheldon2016procedure,sarovar2020detecting,
balewski2025crosstalk}. For source $j$ and victim $k$, the copy retains
the source carrier $\omega_{d,j}$ and has relative phasor
$\beta_{kj}e^{i\theta_{kj}}$. The phase $\theta_{kj}$ includes propagation
phase; the separate \code{delay} parameter shifts the baseband envelope.
The running model sets \code{delay=0} and parameterizes the leakage in each
source--victim direction through amplitude and phase matrices, with victims
indexing rows and sources indexing columns:
\begin{lstlisting}
beta = np.array([[1.0, b12],
                 [b21, 1.0]])
theta = np.array([[0.0, th12],
                  [th21, 0.0]])
eq = ControlEquipment([xy1, xy2])
eq.set_crosstalk_matrix(beta, theta)
\end{lstlisting}
The scheduled tones, their transformation by the control chain, and the
resulting drive Hamiltonians are
\begin{equation}\label{eq:drive}
\begin{aligned}
  s_\ell(t)
    &= \Omega_\ell(t)\cos(\omega_{d,\ell}t+\phi_\ell), \\
  \varepsilon_k(t)
    &= s_k(t) + \beta_{kj}\Omega_j(t)
       \cos\!\bigl(\omega_{d,j}t+\phi_j+\theta_{kj}\bigr), \\
  H_{d,k}(t)
    &= \varepsilon_k(t)\,i(\hat a_k-\hat a_k^\dagger),
\end{aligned}
\end{equation}
where $j$ is the source line and $k$ the victim. The control terms then
complete the lab-frame Hamiltonian:
\begin{equation}\label{eq:htot-full}
  H_{\mathrm{lab}}(t) = H_{\mathrm{static}}
  + \drvterm{\sum_{k=1}^{2} H_{d,k}(t)}
\end{equation}
The resonators are undriven and carry no lines.
The completed device, coupling, and wiring declaration is shown in
Fig.~\ref{fig:anatomy}.

\subsection{Assembling the Chip}\label{sec:assembly}

The devices, couplings, and control equipment are assembled into a
\code{Chip}:
\begin{lstlisting}
chip = Chip([q1, q2, bus, r1, r2],
            couplings=couplings,
            control_equipment=eq,
            frame="rotating", rwa=True)
\end{lstlisting}
\code{frame="rotating"} uses each device's dressed transition as its
reference, while \code{rwa=True} requests removal of counter-rotating
bands. The chip only records these choices; the engine applies them and
records the discarded bands
(Sections~\ref{sec:frames} and~\ref{sec:rwa}).

\subsection{Running an Experiment}\label{sec:running}

\noindent\begin{minipage}{\columnwidth}
Experiments run through a \code{QuantumSequence} on the assembled
chip. Each \code{schedule} call places one pulse on a line, supplying
the envelope, frequency, and phase of Eq.~\eqref{eq:drive};
\code{simulate} solves the scheduled problem over a time grid in ns:
\begin{lstlisting}
seq = QuantumSequence(chip)
seq.schedule(xy1,
    envelope=Gaussian(duration=t_g,
                      amplitude=a_g, sigmas=4.0),
    freq=chip.freq(q1))
res = seq.simulate(tlist,
    initial_state=chip.state({q1: 0}),
    e_ops={q1: q1.projector(1, 1)})
\end{lstlisting}
\end{minipage}
The sequence begins in the dressed ground state
$\lvert\widetilde{0\cdots0}\rangle$, applies a Gaussian pulse to $q_1$ at
its dressed $0\to1$ transition frequency, and records its excited-state
population $\langle\hat P_{1,q_1}\rangle$. \code{chip.freq(q1)} returns
the transition to the dressed eigenstate associated with one excitation of
$q_1$, while \code{chip.state(\{q1: 0\})} returns
$\lvert\widetilde{0\cdots0}\rangle$. The dressed and bare frequencies are
both accessible through the API as \code{chip.freq(q1)} and
\code{q1.freq}, respectively. Because the running model declares no noise,
this call integrates the Schr\"odinger equation; declaring a noise channel
instead selects the Lindblad master equation
(Section~\ref{sec:dissipation}).

The chip declaration can be serialized and reconstructed for reuse in
later simulations:
\begin{lstlisting}
d = chip.to_dict()        # JSON-safe
twin = Chip.from_dict(d)
\end{lstlisting}

Section~\ref{sec:engine} follows this declaration through frame resolution,
Hamiltonian assembly, simulation, sweeps, and gradients.

\begin{figure*}[t]
  \centering
  \includegraphics[width=\textwidth]{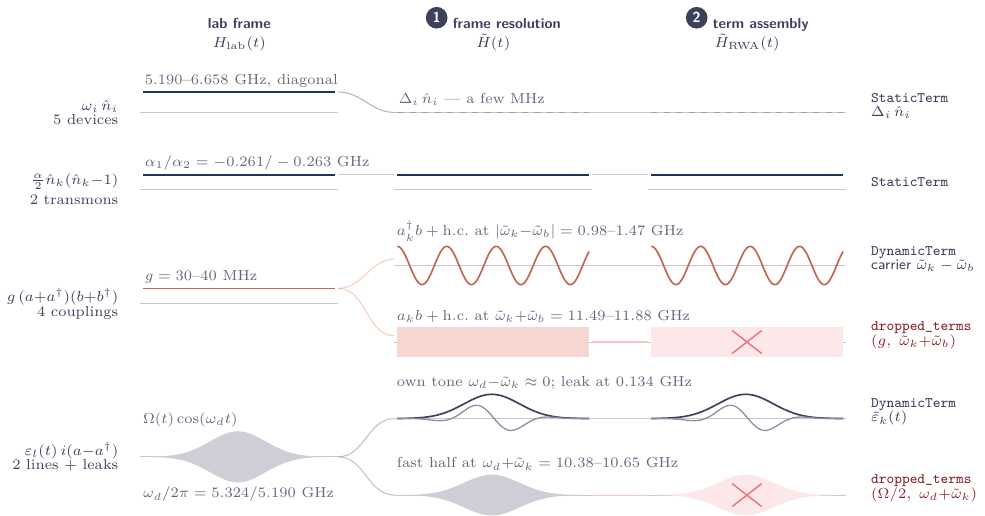}
  \caption{Stages 1 and 2 for the running chip. Rows group the device,
  coupling, and drive terms of Eq.~\eqref{eq:htot-full}; columns show the
  declared lab-frame model, resolved frame, and assembled solver terms.
  Frame resolution leaves residual device detunings; assembly retains the
  difference-frequency coupling and drive bands and records the
  sum-frequency bands in \code{dropped\_terms}. Filled bands denote
  frequencies too high to draw to scale; curves denote slow content.}
  \label{fig:engine-score}
\end{figure*}

\begin{figure*}[t]
  \centering
  \begin{minipage}[t]{0.54\textwidth}
    {\sffamily\fontsize{8}{9.6}\selectfont (a)}\par
    \centering
    \includegraphics[width=\linewidth]{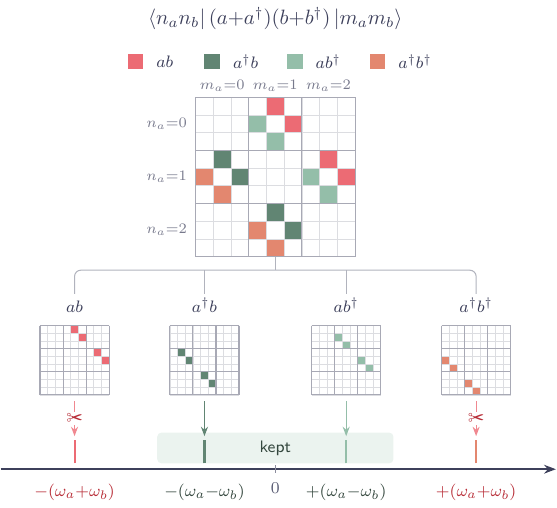}
  \end{minipage}\hfill
  \begin{minipage}[t]{0.42\textwidth}
    {\sffamily\fontsize{8}{9.6}\selectfont (b)}\par
    \centering
    \includegraphics[width=\linewidth]{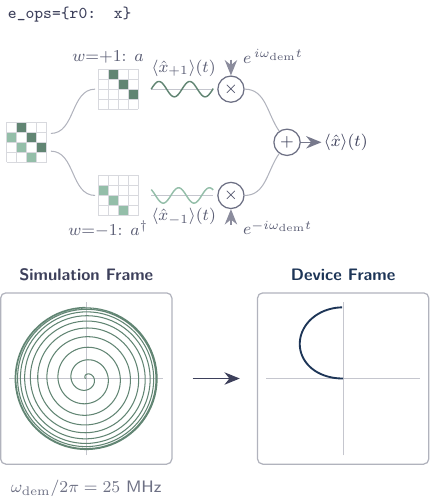}
  \end{minipage}
  \caption{Band decomposition in assembly and readout. (a) Matrix
  elements of a capacitive coupling $(a+a^\dagger)(b+b^\dagger)$
  separate by excitation change into difference-frequency bands retained
  by the RWA and sum-frequency bands that are dropped and recorded.
  (b) A requested quadrature separates into its $w=\pm1$ band matrices
  and is recombined at the demodulation frequency, returning the
  observable in the device's reference frame.}
  \label{fig:bands}
\end{figure*}

\section{The Engine}\label{sec:engine}

In the lab frame, the running Hamiltonian $H_{\mathrm{lab}}(t)$ of
Eq.~\eqref{eq:htot-full} oscillates at the qubit frequencies near
$5.3$~GHz, while the couplings and drive rates that carry the physics
sit at tens of MHz; a solver integrating it must step through every
fast oscillation to capture dynamics that are three orders of
magnitude slower. Rotating each mode at a well-chosen frequency
cancels the fast oscillations, and the rotating-wave approximation
drops what still oscillates at sums of frequencies; done by hand, the
result holds for one model structure, and a new device, coupling, or
drive starts it over. The engine automates this derivation in four
stages (frame resolution, term assembly, observable preparation,
problem packing), and this section follows the concepts they
implement: the frame every term rotates in
(Section~\ref{sec:frames}), the bands the RWA keeps and the observables
read against them (Section~\ref{sec:rwa}), the dissipation the model
admits (Section~\ref{sec:dissipation}), and the frozen problem a backend
compiles into its own solver form (Section~\ref{sec:packing}).
The first two stages are traced term-by-term in
Fig.~\ref{fig:engine-score}.

\subsection{Frame Resolution}\label{sec:frames}

A rotating frame removes a chosen reference oscillation from each mode.
For device $j$, the transformation generated by $\omega_{f,j}\hat n_j$
replaces the term $\omega_j\hat n_j$ by
$(\omega_j-\omega_{f,j})\hat n_j$ and attaches the phase
$e^{-i\omega_{f,j}t}$ to its lowering operator $\hat a_j$. Because the
number operators act on different modes and commute, each frame frequency
can be chosen independently. For the running model, we set
$\omega_{f,j}=\tilde\omega_j$, the dressed $0\to1$ frequency assigned to
mode $j$. Its remaining static term is therefore $\Delta_j\hat n_j$, where
$\Delta_j=\omega_j-\tilde\omega_j$ is the difference between the declared
bare frequency and the dressed reference. Couplings and drives do not
disappear under this transformation; their ladder operators acquire phases
that expose slow difference-frequency terms and fast sum-frequency terms:
\begin{equation}\label{eq:htot-rotating}
\begin{split}
  \tilde H(t) ={}& \devterm{\sum_j \Delta_j \hat n_j} + \sum_{k=1}^{2} \Bigl[
    \devterm{\tfrac{\alpha}{2}\hat n_k(\hat n_k - 1)} \\
    &+ \cplterm{g_{kb}\bigl( \hat a_k^\dagger \hat b\, e^{i(\tilde\omega_k-\tilde\omega_b)t}
       + \hat a_k \hat b\, e^{-i(\tilde\omega_k+\tilde\omega_b)t}} \\
    &\qquad\quad {}\cplterm{+ \mathrm{h.c.} \bigr)} \\
    &+ \cplterm{g_{kr}\bigl( \hat a_k^\dagger \hat c_k\, e^{i(\tilde\omega_k-\tilde\omega_{r_k})t}
       + \hat a_k \hat c_k\, e^{-i(\tilde\omega_k+\tilde\omega_{r_k})t}} \\
    &\qquad\quad {}\cplterm{+ \mathrm{h.c.} \bigr)} \\
    &+ \drvterm{\varepsilon_k(t)\, i\bigl( \hat a_k e^{-i\tilde\omega_k t}
       - \hat a_k^\dagger e^{i\tilde\omega_k t} \bigr)} \Bigr]
\end{split}
\end{equation}
with $j$ running over all five devices. The two coupling terms describe
different processes. The exchange term $\hat a_k^\dagger\hat b$ transfers
one excitation between the modes and rotates at the difference of their
frame frequencies. The pair term $\hat a_k\hat b$, together with its
Hermitian conjugate, annihilates or creates two excitations and rotates at
the sum of those frequencies. Each drive similarly separates into
components at $\omega_d-\tilde\omega_k$ and
$\omega_d+\tilde\omega_k$.

Stage 1 resolves the frame declaration into a \code{ResolvedFrame}
holding one frequency per device. The \code{frame="rotating"} choice
of Section~\ref{sec:assembly} assigns each device its dressed $0 \to 1$
frequency; \code{"lab"} assigns $\omega_{f,j}=0$, where zero is the
frame's rotation frequency rather than a bare or dressed device
transition. A scalar or per-device mapping assigns the frequencies
directly. No term is dropped at this stage.

\subsection{Band Decomposition: RWA and Observables}\label{sec:rwa}

The RWA keeps a Hamiltonian's slow terms and
discards its fast ones, and which is which follows from how each term
rotates in the resolved frame. Every operator decomposes into
\emph{bands} by excitation change: the matrix element
$\langle n|O|m\rangle$ belongs to the band of weight $w = m - n$, which
rotates at $w\,\tilde\omega$, and a two-body band
$(\Delta a, \Delta b)$ rotates at
$\Delta a\,\tilde\omega_a + \Delta b\,\tilde\omega_b$. A coupling's
excitation-conserving bands, $\Delta a + \Delta b = 0$, rotate at the
difference of the frame frequencies and stay near-resonant; the rest
rotate at their sum and oscillate fast. A drive splits the same way,
into a component at $\omega_d - \tilde\omega_k$ and one at
$\omega_d + \tilde\omega_k$. Assembly and observable readout use the
same band decomposition: assembly splits each operator into bands and
drops the fast ones [Fig.~\ref{fig:bands}(a)]~\cite{jaynes1963,gambetta2006,rigetti2010,magesan2020},
while observable readout recombines them, each band at its own rate
[Fig.~\ref{fig:bands}(b)].

Stage 2 performs the decomposition for every operator and, under
\code{rwa=True}, drops the fast bands: a coupling keeps its
$\Delta a + \Delta b = 0$ bands by default, although a coupling
class may declare a different retained set, and each drive keeps its
difference component.

Applied to Eq.~\eqref{eq:htot-rotating}: every pair term is dropped,
every exchange term survives at its detuning, and each drive keeps
its difference component:
\begin{equation}\label{eq:htot-rwa}
\begin{split}
  \tilde H_{\mathrm{RWA}}(t) ={}& \devterm{\sum_j \Delta_j \hat n_j}
    + \sum_{k=1}^{2} \Bigl[
    \devterm{\tfrac{\alpha}{2}\hat n_k(\hat n_k - 1)} \\
    &+ \cplterm{g_{kb}\bigl( \hat a_k^\dagger \hat b\, e^{i(\tilde\omega_k-\tilde\omega_b)t}
      + \mathrm{h.c.} \bigr)} \\
    &+ \cplterm{g_{kr}\bigl( \hat a_k^\dagger \hat c_k\, e^{i(\tilde\omega_k-\tilde\omega_{r_k})t}
      + \mathrm{h.c.} \bigr)} \\
    &+ \drvterm{i\bigl( \tilde\varepsilon_k(t)\, \hat a_k
      - \tilde\varepsilon_k^*(t)\, \hat a_k^\dagger \bigr)} \Bigr]
\end{split}
\end{equation}
with $\tilde\varepsilon_k$ the retained component of the line's
signal: for the qubit's own tone
$\tfrac{1}{2}\Omega_k(t)\, e^{i((\omega_{d,k}-\tilde\omega_k)t+\phi_k)}$,
and for the leaked copy of the other line a tone detuned by
$\omega_{d,j} - \tilde\omega_k$, the crosstalk the study of
Section~\ref{sec:models} measures.

Each dropped band is recorded on the assembled problem with its
amplitude and oscillation frequency
(\code{problem.hamiltonian.dropped\_terms}), and a band of matrix
element $\lambda$ oscillating at $\omega$ shifts the levels it couples
by $\lambda^2/\omega$ at second order~\cite{bloch1940,zueco2009}.
The resulting static frequency shift for a coupling and accumulated
phase for a pulsed drive agree with second-order perturbation theory
(App.~\ref{app:rwa-validity}). The stage's output is a set of
\code{StaticTerm}s and \code{DynamicTerm}s over backend-neutral
\code{CanonicalOperator}s (dense, sparse, or diagonal,
preserved as declared), with the collapse operators assembled
alongside them (Section~\ref{sec:dissipation}).

Stage 3 prepares the requested observables, keyed by the objects
themselves:
\begin{lstlisting}
e_ops = {q1: q1.projector(1, 1),
         q2: q2.projector(1, 1)}
\end{lstlisting}
Each requested observable decomposes into the same excitation-change
bands, which enter the solver separately. After the solve, their
expectations are recombined with a phase at the device's demodulation
frequency (Fig.~\ref{fig:bands}(b)), the
difference between its reporting reference and the integration frame,
$\omega_{\mathrm{dem}} = \omega_{\mathrm{ref}} - \tilde\omega$:
\begin{equation}\label{eq:demod}
  \langle \hat O \rangle(t) = \sum_w e^{\,i w\, \omega_{\mathrm{dem}}
  t}\, \langle \hat O_w \rangle(t)
\end{equation}
the inverse of the rotation applied to the Hamiltonian. Results
therefore return in each device's reporting frame, corresponding to the
slowly varying signal obtained after laboratory demodulation, regardless
of the frame in which the solver performs the simulation.
Populations live in the $w = 0$ band and carry no phase; a transverse
observable such as $\langle a \rangle$ returns as its slow envelope
rather than a 5-GHz oscillation. Our chip integrates in the reference
frame itself; $\omega_{\mathrm{dem}} = 0$, and the correction is
therefore the identity.

\subsection{Dissipation}\label{sec:dissipation}

Our chip so far is closed: its state is a vector $|\psi\rangle$
evolving under the Schr\"odinger equation~\cite{schrodinger1926}:
\begin{equation}\label{eq:schrodinger}
  i\,|\dot\psi\rangle = \tilde H_{\mathrm{RWA}}(t)\,|\psi\rangle
\end{equation}
A device coupled to its environment is an open system: its state is a
density matrix $\rho$ evolving under the Lindblad master
equation~\cite{lindblad1976}:
\begin{equation}\label{eq:lindblad}
  \dot\rho = -i\,[\tilde H_{\mathrm{RWA}}(t), \rho]
  + \sum_j \Bigl( \hat L_j \rho \hat L_j^\dagger
  - \tfrac{1}{2}\bigl\{\hat L_j^\dagger \hat L_j, \rho\bigr\} \Bigr)
\end{equation}
where each dissipation channel defines a collapse operator $\hat L_j$.
Each collapse operator corresponds to a declared dissipation channel; none
is required. Built-in devices provide common optional parameterizations.
When a \code{DuffingTransmon} is assigned a relaxation time $T_1$, it adds
amplitude damping $\sqrt{(\bar n{+}1)/T_1}\,\hat a$ and
$\sqrt{\bar n/T_1}\,\hat a^\dagger$ ($\bar n$ the bath occupation, zero
by default). Assigning a coherence time $T_2$ adds pure dephasing
$\sqrt{2\gamma_\varphi}\,\hat n$ with the rate~\cite{krantz2019guide}:
\begin{equation}\label{eq:gamma-phi}
  \gamma_\varphi = \frac{1}{T_2} - \frac{1}{2T_1}
\end{equation}
A \code{Resonator} assigned a quality factor $Q$ adds photon loss
$\sqrt{\kappa}\,\hat c$ with
$\kappa = \omega_r/Q$~\cite{blais2021circuit}. Leaving any of these fields
unset omits that channel. A model may therefore leave a qubit's intrinsic
$T_1$ and $T_2$ unset while assigning a finite $Q$ to its readout
resonator, so qubit relaxation arises through its coupling to the lossy
resonator rather than from an imposed intrinsic lifetime.

These times are declared by writing them onto the same device objects
of Section~\ref{sec:devices},
\begin{lstlisting}
q1.T1, q1.T2 = T1_q, T2_q
q2.T1, q2.T2 = T1_q, T2_q
r1.quality_factor = Q_r
r2.quality_factor = Q_r
\end{lstlisting}
The same parameters can instead be set in one replace-all call,
\code{chip.set\_noise(\{q1: dict(T1=T1\_q, T2=T2\_q), ...\})}. These
assignments define an open-system variant; they are not made in the
running study. Its two qubits, bus, and readout resonators are all
lossless, so no collapse operators enter its simulations. The engine
tracks later noise assignments and rebuilds the collapse operators on
the next solve. Shared environments are represented by chip-owned
\code{Bath} objects. Despite being attached at the chip level, a bath need
not act on the entire chip: \code{targets} may specify any subset of
devices, while omitting it applies the bath to every device. Available
recipes include collective decay, correlated dephasing, and thermal baths.

As with devices, couplings, and drives, a new local noise channel is
defined on the device class alone:
\begin{lstlisting}
def flux_dephasing(q):
    xp = get_default_backend().array_module
    return [xp.sqrt(2 * q.gamma_flux)
            * q.number_operator()]

flux_channel = NoiseChannel(
    "flux_dephasing", ("gamma_flux",),
    flux_dephasing)

class FluxNoisyTransmon(DuffingTransmon):
    gamma_flux: Scalar = parameter(
        positive=True, unit="GHz")
    _noise_channels = (
        DuffingTransmon._noise_channels
        + (flux_channel,))
\end{lstlisting}

\subsection{Partitioning, Packing, and Backends}\label{sec:packing}

\begin{figure*}[t]
  \centering
  \includegraphics[width=\textwidth]
    {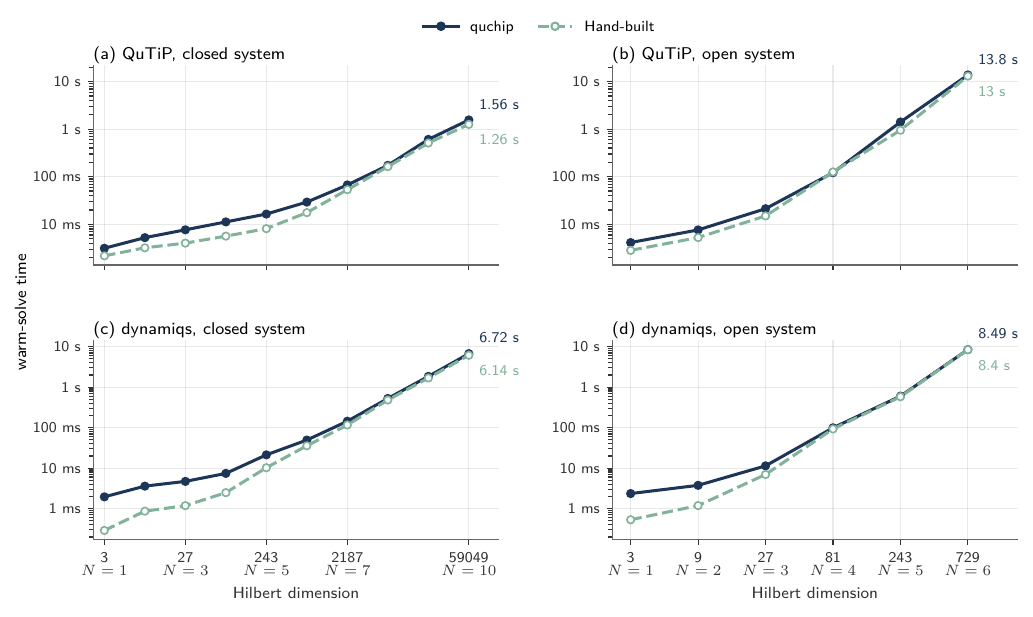}
  \caption{Warm-solve times for driven chains of $N$ three-level
  transmons constructed through quchip and directly in each backend.
  (a) QuTiP, closed system; (b) QuTiP, open system; (c) dynamiqs,
  closed system; and (d) dynamiqs, open system. Closed-system
  simulations reach $d=3^{10}=59049$; open-system simulations reach
  $d=3^6=729$ and propagate $d\times d$ density matrices. Each point is
  the median of five solves on the same 401-point time grid after
  one-time construction and compilation.}
  \label{fig:overhead}
\end{figure*}

\begin{figure*}[t]
  \centering
  \includegraphics[width=\textwidth]
    {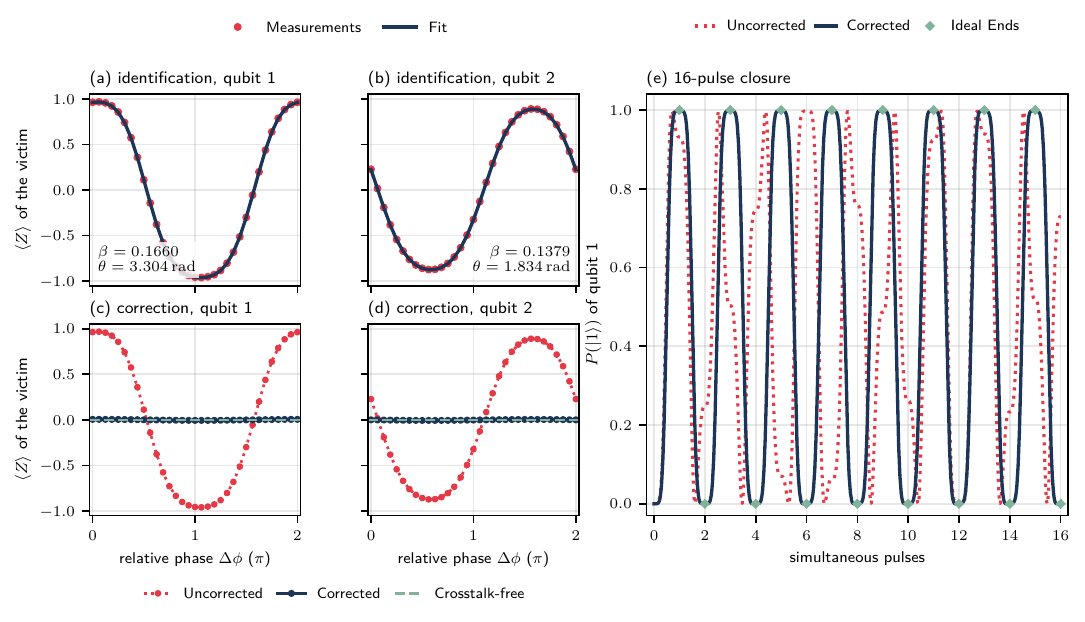}
  \caption{Crosstalk identification and correction on the five-device
  model. (a,b) Simulated phase sweeps and fits of
  Eq.~\eqref{eq:crosstalk-response}. (c,d) Responses before and after
  correction; dashed lines mark the crosstalk-free response. (e) Qubit-1
  population over sixteen simultaneous $\pi$ pulses; the corrected
  evolution follows the ideal pulse-end values. All panels use the same
  full model.}
  \label{fig:crosstalk-loop}
\end{figure*}

\begin{figure*}[t]
  \centering
  \includegraphics[width=\textwidth]
    {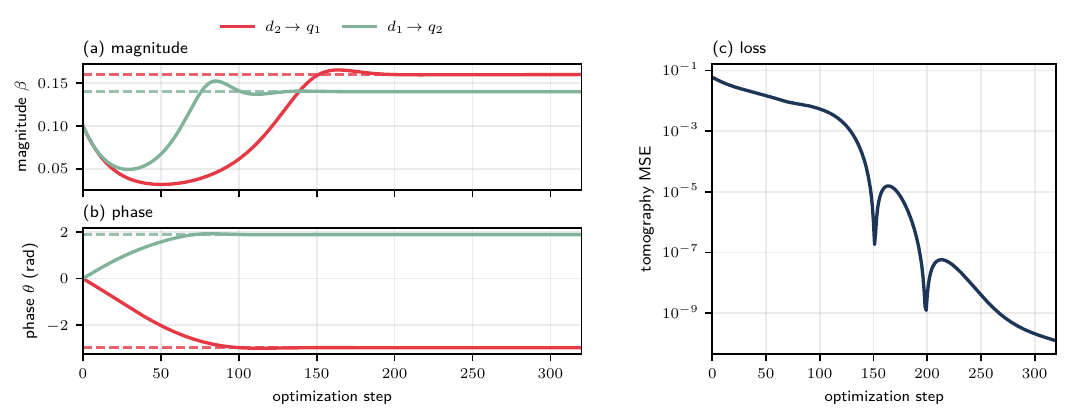}
  \caption{Gradient-based recovery of directed crosstalk from tomography on
  the reduced two-qubit model. (a,b) Inferred magnitudes and phases over
  320 optimization steps; dashed lines mark the injected values, with phases
  shown on the continuous branch from zero. (c) Tomography loss over the
  same optimization.}
  \label{fig:crosstalk-gradient}
\end{figure*}

When two subsets of a chip's devices share no interaction (no coupling,
bath, or drive crosstalk between them), their dynamics are independent
and the joint state remains a tensor product of the two. The engine
partitions the chip into the connected components of its interaction
graph and solves each in its own Hilbert space rather than the joint
product space. Observables read from their own component, and
cross-component correlators are products of local expectations; a joint
state is reconstructed only on request. The partitioning is exact,
agreeing with the joint solve to solver tolerance, and \code{simulate()}
applies it by default (\code{partition=False} forces the joint solve).
The running chip is a single connected component. Reducing it to a smaller
effective model therefore requires an explicit approximation;
\code{eliminate()} provides such a reduction for far-detuned modes by
folding their Lamb shifts, mediated couplings, and, where applicable,
Purcell decay into the surviving devices.

Each component's assembled description is packed into a single frozen
\code{SolveProblem}: the Hamiltonian description, the collapse
operators, the initial state, the time grid, the prepared observables,
the resolved frame, and solver options. The record is immutable, and an
unchanged chip reuses its assembled Hamiltonian rather than rebuilding
it. Assigning or replacing noise parameters changes the model state, so
the next simulation rebuilds the problem and its collapse operators.

A backend converts the packed problem into its solver's native form
and runs it. Its interface provides an array namespace and dense-array
conversion, conversions between \code{CanonicalOperator} and the native
operator type, tensor products and two-body embedding, and two solve
entry points: \code{sesolve} for
Eq.~\eqref{eq:schrodinger} and \code{mesolve} for
Eq.~\eqref{eq:lindblad}.

The two shipped backends make opposite conversion choices from the
same description. The QuTiP backend builds \code{Qobj} operators with
time-dependent coefficient callables; the dynamiqs backend builds JAX
arrays and time-callables and caches its jitted solves. Problem
preparation, state coercion, and batched sweeps are inherited from shared
defaults and overridden only where a native form is faster.

Section~\ref{sec:assembly} selected the backend at chip construction; it
can also be scoped to one call, letting the same chip serve both:
\begin{lstlisting}
seq.simulate(tlist)  # chip backend
seq.simulate(tlist, backend="dynamiqs")  # override
\end{lstlisting}

Both backends expose the same sweep interface: \code{vary} turns a
scheduled quantity into a batch axis, and \code{simulate\_batch}
integrates the resulting batch.
\begin{lstlisting}
h = seq.schedule(ctrl, envelope=..., freq=f)
batch = seq.simulate_batch(
    h.vary("amplitude", amps),
    tlist=tlist,
    e_ops={q: q.projector(1, 1)})
\end{lstlisting}
Any scheduled quantity is a valid axis: an envelope parameter, the
carrier frequency, the phase, or the pulse start time. Several axes
combine as a full grid, or in lockstep through \code{seq.zip}. Dynamiqs
integrates the batch as a single \code{vmap}ped solve, while QuTiP fans
the concrete sweep across a pool of \code{loky} worker processes.

To isolate the cost of quchip's modeling layer, we simulate the same
driven transmon chains through quchip and through Hamiltonians written
directly for QuTiP and dynamiqs. Figures~\ref{fig:overhead}(a) and
\ref{fig:overhead}(c) compare closed-system solves, while
Figs.~\ref{fig:overhead}(b) and \ref{fig:overhead}(d) compare open-system
solves. In every case, the quchip and hand-built timings converge as the
Hilbert dimension grows. quchip therefore retains the scaling of the
underlying solver rather than introducing an additional
dimension-dependent cost.

\subsection{Gradients}\label{sec:diff}

The dynamiqs backend additionally exposes derivatives of simulated
observables with respect to JAX-traced model and pulse parameters. A scalar
loss can therefore be differentiated with respect to static and
time-dependent parameters together. Sweeps are supported by both backends,
but gradients require dynamiqs.
Figure~\ref{fig:pipeline} summarizes the forward simulation and
differentiable paths through the engine.

Section~\ref{sec:models} uses forward simulations to generate the
phase-sweep responses of the fitted five-device chip and test the
resulting crosstalk correction. It then applies \code{eliminate()} to
reduce the chip to two qubits and uses the differentiable path to recover
the four directed crosstalk parameters from simulated tomography.

\section{Closing the Experimental Loop}\label{sec:models}

An experimental loop begins with a measurement of a physical device,
uses the result to create or update a predictive model of the measured
response, and closes when a subsequent measurement tests a prediction
made by that model. We follow the crosstalk-identification procedure of
Balewski et al.~\cite{balewski2025crosstalk}, using its reported line
phasors and a full quchip simulation in place of the physical device.
Phase sweeps identify the line response and give a correction, which we
test on the same five-device model. We then reverse the problem: after
reducing the fitted chip to two qubits, we treat the four directed
crosstalk parameters as unknown and recover them through differentiation
from simulated tomography.
Figures~\ref{fig:crosstalk-loop}(a) and \ref{fig:crosstalk-loop}(b) show
the two identification sweeps, Figs.~\ref{fig:crosstalk-loop}(c) and
\ref{fig:crosstalk-loop}(d) show the corresponding corrections, and
Fig.~\ref{fig:crosstalk-loop}(e) tests the corrected response over repeated
simultaneous pulses.

An experimental chip is brought into quchip by fitting its model to
measured dressed quantities rather than assigning them directly as bare
parameters. Here, the declaration in Section~\ref{sec:building} supplies the
five-device structure and starting parameters. Approximate dressed transition
frequencies, anharmonicities, and readout frequencies for Q1 and Q2 are
digitized from the device-characterization figure of Hashim et
al.~\cite{hashim2025midcircuit}; the plot resolution is about $1$--$2$ MHz.
The bus frequency and coupling strengths remain representative because those
values are not reported. A dressed-to-bare fit, performed with
\code{fit\_a\_dress()}, adjusts the bare parameters to reproduce the dressed
targets while holding the four coupling strengths fixed.
The fitted chip is held fixed throughout crosstalk identification and
correction, then reduced for the inverse calculation at the end of the section.

To measure leakage from control line $j$ onto qubit $k$, the protocol
applies simultaneous equal-area pulses: one through the intended line
and another through line $j$, both with the frequency of qubit $k$ and
with relative phase $\Delta\phi$. In the two-level model, the intended
and leaked fields add coherently:
\begin{equation}\label{eq:crosstalk-response}
\begin{aligned}
  \eta(\Delta\phi)
  &= \sqrt{1+\beta_{kj}^{2}
      +2\beta_{kj}\cos(\Delta\phi-\theta_{kj})},
  \\
  \langle Z_k\rangle
  &= \cos\!\left[\vartheta_k\eta(\Delta\phi)\right],
\end{aligned}
\end{equation}
where $\vartheta_k$ is the rotation produced by the intended pulse alone. Sweeping
$\Delta\phi$ determines the leakage amplitude $\beta_{kj}$ and phase
$\theta_{kj}$; reversing the two lines determines $\beta_{jk}$ and
$\theta_{jk}$.

We apply the protocol to the five-device model, assigning the control
equipment the reported line phasors $0.16e^{i3.31}$ and
$0.14e^{i1.885}$. Each phase-sweep point is obtained from the complete
driven Hamiltonian. We determine the intended-line rotation from an
independent isolated amplitude sweep, then fit only $\beta$ and $\theta$ in
Eq.~\eqref{eq:crosstalk-response}; the maximum RMS residual is $0.00272$.
The fitted phasors include both control-line leakage and coupling through
the dressed transitions [Figs.~\ref{fig:crosstalk-loop}(a) and
\ref{fig:crosstalk-loop}(b)].

The phase sweeps determine the complex relative response matrix
\begin{equation}\label{eq:crosstalk-matrix}
\widehat C=
\begin{pmatrix}
1 & \hat\beta_{12}e^{i\hat\theta_{12}}\\
\hat\beta_{21}e^{i\hat\theta_{21}} & 1
\end{pmatrix}.
\end{equation}
The complex pulse envelopes $a_1$ and $a_2$ that would be applied
without crosstalk are converted into the programmed envelopes $u_1$
and $u_2$ according to
\begin{equation}\label{eq:crosstalk-correction}
\begin{pmatrix}u_1\\u_2\end{pmatrix}
=
\widehat C^{-1}
\begin{pmatrix}a_1\\a_2\end{pmatrix}.
\end{equation}
Two isolated rotations normalize the columns of the inverse before it is
applied; this fixes the single-line amplitude calibration implicit in
$a_1$ and $a_2$ without changing the fitted off-diagonal phasors.
In quchip, the same relative response matrix $\widehat C$ is calculated
independently from the fitted chip's dressed drive matrix elements and
declared line crosstalk; its complex entries differ from the phase-sweep fit
by at most $0.00134$ (Appendix~\ref{app:dressed-crosstalk}).
Applied to the unchanged five-device model, the correction suppresses
the effective leakage by more than two orders of magnitude
[Figs.~\ref{fig:crosstalk-loop}(c) and \ref{fig:crosstalk-loop}(d)]. Over
sixteen simultaneous $\pi$ pulses, the corrected population at each pulse
end differs from the ideal value by less than $0.3$ percentage points
[Fig.~\ref{fig:crosstalk-loop}(e)].

To test the differentiable path, we first reduce the fitted five-device chip by
applying \code{eliminate()} to adiabatically remove the two readout resonators
and the bus~\cite{bravyi2011schrieffer}. This reduces the Hilbert-space dimension from 576 to 16, leaving two
qubits connected by one effective capacitive interaction. We then infer
$\boldsymbol{\lambda}=(\beta_{12},\theta_{12},\beta_{21},\theta_{21})$
from ordinary tomography on the reduced chip. Two directional weak probes
produce four final $X$ and $Y$ expectations, collected in
$\mathbf y(\boldsymbol{\lambda})$. Given the fixed tomography data
$\mathbf y^{\mathrm{data}}$, we minimize:
\begin{equation}\label{eq:crosstalk-tomography-loss}
  \mathcal L(\boldsymbol{\lambda})
  = \frac{1}{4}\left\|
      \mathbf y(\boldsymbol{\lambda})-\mathbf y^{\mathrm{data}}
    \right\|_2^2
\end{equation}
Starting from $\beta=0.10$ and $\theta=0$ in both directions, Adam
steps~\cite{kingma2014adam} along gradients through the control
signal chain, Hamiltonian construction, and time evolution reduce the
tomography loss [Fig.~\ref{fig:crosstalk-gradient}(c)] and recover the
injected magnitudes and phases [Figs.~\ref{fig:crosstalk-gradient}(a) and
\ref{fig:crosstalk-gradient}(b)], with a maximum complex error of
$1.5\times10^{-5}$.

\section{Conclusion}\label{sec:conclusion}

We introduced quchip to make a quantum-chip model reusable across simulation
tasks. A device carries its local physics, a coupling supplies an interaction,
and a drive connects a classical control signal to a device. quchip assembles
these pieces when a simulation is prepared, resolving the chosen frame and
approximations before passing the problem to QuTiP or dynamiqs. The chip
description is therefore not tied to one solver or one calculation: it can be
extended, reduced, swept, or differentiated without deriving a new Hamiltonian
for each task.

On the five-device chip, fitting bare parameters to dressed observables fixed
the model before phase sweeps identified the two directed line leaks,
inversion of the response matrix supplied a correction, and repeated pulses
tested it. Applying \code{eliminate()} to the two readout resonators and bus
then reduced the model from 576 to 16 dimensions; gradients through the
reduced simulation recovered the four crosstalk parameters from ordinary
tomography. These results demonstrate both directions of the modeling loop
on one physical declaration: prediction and correction on the full chip,
followed by parameter recovery on its controlled effective model.

\acknowledgments{I first thank UTA for invaluable discussions, thoughtful
feedback on the manuscript, encouragement, and enduring friendship. I am
grateful to Dr.\ Ahmed Hajr for his guidance and mentorship, and to Albaraa
Shafi for his discussions, advice, and support. I also thank Abdullah
Alqarni, Abdullah Alnafisah, Ryan Alghuraybi, and Sufyan Alshaalan for their
helpful conversations and encouragement. Above all, I am grateful to my
fiancée for her constant love, support, and unwavering encouragement, and for
believing in me throughout the development of this project.}

\bibliographystyle{unsrtnat}
\bibliography{quchip}

\onecolumngrid
\appendix
\section{Dressed Crosstalk Response}\label{app:dressed-crosstalk}

The phase-sweep coefficients combine classical leakage between control
lines with cross-driving through the dressed transitions. Let
$C^{\mathrm{line}}_{\ell j}$ map programmed line $j$ onto physical line
$\ell$, and let $\hat D_\ell$ be the corresponding drive operator. The
calculated response is:
\begin{equation}\label{eq:dressed-crosstalk}
\begin{aligned}
  D_{k\ell}
  &= \langle\widetilde{1_k}|\hat D_\ell|\widetilde{0}\rangle
  \\
  R
  &= D C^{\mathrm{line}}
  \\
  C^{\mathrm{eff}}_{kj}
  &= \frac{R_{kj}}{R_{kk}}
\end{aligned}
\end{equation}
The row normalization gives $C^{\mathrm{eff}}$ a unit diagonal. On the
fitted five-device model, its off-diagonal entries agree with those of
$\widehat C$ obtained from the phase sweeps within one percent.

\section{Rotating-Wave Validity}\label{app:rwa-validity}

The terms discarded by the RWA remain available
through \code{problem.hamiltonian.dropped\_terms}.
Table~\ref{tab:rwa-validity} compares their effect with a full calculation in
which \code{rwa=False}: a counter-rotating
coupling produces a static frequency shift, whereas a counter-rotating
drive accumulates a phase over the pulse.

\begin{table}[h!]
  \centering\footnotesize
  \begin{tabular}{@{}lllll@{}}
    \toprule
    Test & Observable and second-order prediction & Full model & Prediction & Relative error \\
    \midrule
    pair, $g/2\pi=10$ MHz &
      $\delta\omega/2\pi=g^2/[2\pi(\omega_a+\omega_b)]$ (GHz) &
      $9.90098\times10^{-6}$ & $9.90099\times10^{-6}$ & $10^{-6}$ \\
    pair, $g/2\pi=20$ MHz &
      $\delta\omega/2\pi$ (GHz) &
      $3.96038\times10^{-5}$ & $3.96040\times10^{-5}$ & $4\times10^{-6}$ \\
    pair ratio, $20/10$ &
      $\delta\omega_{20}/\delta\omega_{10}$ &
      $3.99999$ & $4$ & $3\times10^{-6}$ \\
    drive, $\Omega/2\pi=0.1$ GHz &
      $\Delta\phi=-\int_0^T\!\Omega(t)^2/[2(\omega_q+\omega_d)]\,dt$ (rad) &
      $-2.3110\times10^{-2}$ & $-2.3201\times10^{-2}$ & $0.4\%$ \\
    drive, $\Omega/2\pi=0.2$ GHz &
      $\Delta\phi$ (rad) &
      $-9.1372\times10^{-2}$ & $-9.2805\times10^{-2}$ & $1.5\%$ \\
    drive ratio, $0.2/0.1$ &
      $\Delta\phi_{0.2}/\Delta\phi_{0.1}$ &
      $3.9538$ & $4$ & $1.2\%$ \\
    \bottomrule
  \end{tabular}
  \caption{Effects of discarded counter-rotating bands compared with
  second-order predictions. The coupling test uses a 5.0/5.1-GHz pair and
  exact diagonalization; the drive test uses a 5.0-GHz qubit, a 30-ns
  Gaussian pulse, and a 4.0-GHz carrier. Ratio rows test the predicted
  quadratic scaling.}
  \label{tab:rwa-validity}
\end{table}

\end{document}